\documentclass[%
 reprint,
 superscriptaddress,
 longbibliography,
 amsmath,amssymb,
 aps,
 prmaterials,
 floatfix,
]{revtex4-2}

\usepackage{graphicx}
\usepackage{dcolumn}
\usepackage{bm}
\usepackage[hidelinks]{hyperref}
\usepackage{siunitx}
\usepackage[version=4]{mhchem}
\usepackage{xcolor}

\newcommand{\den}[1]{\SI{#1}{\per\square\centi\meter}}
\newcommand{\mob}[1]{\SI[per-mode=symbol]{#1}{\square\centi\meter\per\volt\per\second}}
\newcommand{\mobr}[2]{\SIrange[per-mode=symbol,range-units=single]{#1}{#2}{\square\centi\meter\per\volt\per\second}}
\DeclareSIUnit\sq{\ensuremath{\Box}}

\begin{document}

\preprint{xxxxxx}

\title{Repairing the Surface of InAs-based Topological Heterostructures}

\author{S. J. Pauka}
    \email{sebastian.pauka@sydney.edu.au}
\author{J. D. S. Witt}
    \affiliation{ARC Centre of Excellence for Engineered Quantum Systems, School of Physics, The University of Sydney, Sydney, NSW 2006, Australia.}
\author{C. N. Allen}
    \affiliation{Microsoft Quantum Sydney, The University of Sydney, Sydney, NSW 2006, Australia.}
\author{B. Harlech-Jones}
\author{A. Jouan}
    \affiliation{ARC Centre of Excellence for Engineered Quantum Systems, School of Physics, The University of Sydney, Sydney, NSW 2006, Australia.}
\author{G. C. Gardner}
\author{S. Gronin}
\author{T. Wang}
\author{C. Thomas}
    \affiliation{Birck Nanotechnology Center, Purdue University, West Lafayette, IN 47907, USA.}
    \affiliation{Microsoft Quantum Purdue, Purdue University, West Lafayette, IN 47907, USA.}
\author{M. J. Manfra}
    \affiliation{Department of Physics and Astronomy, Purdue University, West Lafayette, IN 47907, USA.}
    \affiliation{Birck Nanotechnology Center, Purdue University, West Lafayette, IN 47907, USA.}
    \affiliation{Microsoft Quantum Purdue, Purdue University, West Lafayette, IN 47907, USA.}
    \affiliation{School of Materials Engineering and School of Electrical and Computer Engineering, Purdue University, West Lafayette, IN 47907, USA.}
\author{D. J. Reilly}
    \affiliation{ARC Centre of Excellence for Engineered Quantum Systems, School of Physics, The University of Sydney, Sydney, NSW 2006, Australia.}
    \affiliation{Microsoft Quantum Sydney, The University of Sydney, Sydney, NSW 2006, Australia.}
\author{M. C. Cassidy}
    \affiliation{Microsoft Quantum Sydney, The University of Sydney, Sydney, NSW 2006, Australia.}

\date{\today}

\begin{abstract}
Candidate systems for topologically-protected qubits include two-dimensional electron gases (2DEGs) based on heterostructures exhibiting a strong spin-orbit interaction (SOI) and superconductivity via the proximity effect. For InAs- or InSb-based materials, the need to form shallow quantum wells to create a hard-gapped $p$-wave superconducting state often subjects them to fabrication-induced damage, limiting their mobility. Here we examine scattering mechanisms in processed InAs 2DEG quantum wells and demonstrate a means of increasing their mobility via repairing the semiconductor-dielectric interface. Passivation of charged impurity states with an argon-hydrogen plasma results in a significant increase in the measured mobility and reduction in its variance relative to untreated samples, up to \mob{45300} in a \SI{10}{\nano\meter} deep quantum well.
\end{abstract}

\maketitle


\section{\label{sec:intro}Introduction}

Interest in proximitized InAs- and InSb-2DEGs has intensified recently due to their potential application in spintronics \cite{spintronics} and topological quantum computation \cite{PhysRevLett.105.077001,s41578-018-0003-1}. These materials can exhibit superconductivity via the proximity effect, induced by the presence of aluminum deposited on their surface, which strongly couples to the quantum well. The induced superconductivity combined with strong SOI (and a large Land\'e g-factor) results in the formation of Majorana zero modes (MZMs), now observed in both nanowires \cite{Mourik1003,AlbrechtNature} and 2DEGs \cite{PhysRevLett.119.136803,PhysRevLett.119.176805}, at the boundaries of the topological superconductor \cite{RevModPhys.83.1057,Kitaev_2001,doi:10.1146/030212-184337}. Interest in MZMs, which are emergent quasi-particles hypothesized to have non-abelian exchange statistics, stems from their potential to provide topological protection to quantum information \cite{RevModPhys.80.1083}.

Early experimental platforms for realizing MZMs in both nanowires and 2DEGs utilized superconductors deposited ex-situ, however these systems demonstrated a a significant density of states sub-gap that obscured the signatures of the MZM \cite{PhysRevB.88.064506,PhysRevLett.110.186803}. Alternatively, in-situ deposition of a superconductor, such as epitaxially growing aluminum directly after semiconductor growth, results in a significant improvement in the quality of the superconducting gap \cite{nnano.2014.306,hard_gap_2deg}, and the realization of  a quantized zero-bias peak \cite{nature26142}. In-situ deposition poses additional fabrication challenges however, as the Al must be removed to define the topological region of the device. Removal via a wet-etch solution selective to Al is a highly exothermic reaction that results in damage to the surface of the semiconductor. This damage manifests as increased roughness and induced impurities, lowering the mobility of the 2DEG \footnote{From \mob{44000} \cite{shabani_transport} down to \mobr{1000}{2000} \cite{hard_gap_2deg,PhysRevLett.115.127001}} and compromising the fragile induced $p$-wave superconducting pairing \cite{PhysRevB.83.184520,PhysRevB.85.140513}. Further, since the length scale over which hard-gap superconductivity is maintained across a clean interface is set by the height and thickness of the barrier \cite{PhysRevLett.110.186803} burying the 2DEG deep in the heterostructure is not feasible \cite{PhysRevB.93.155402}. Taken together these aspects point to a need to develop new fabrication techniques that maintain or repair defects introduced at the surface.

Here we investigate the scattering mechanisms that reduce mobility in shallow InAlAs/InAs/InGaAs 2DEG heterostructures following wet-etch of the proximitizing superconductor. By studying Hall mobility as a function of density, we show that surface scattering is the dominant mechanism for reduced mobility in shallow 2DEG samples. We demonstrate that the mobility can be increased and the variance of mobility reduced by exposing the sample to an in-situ Ar-H plasma, prior to deposition of a protective ALD-grown \ce{Al2O3} coating. We compare the mobility of these to control samples exposed to an in-situ trimethylaluminum (TMA) pre-treatment, or to samples without any pre-treatment, prior to oxide growth.

\begin{figure}
\includegraphics{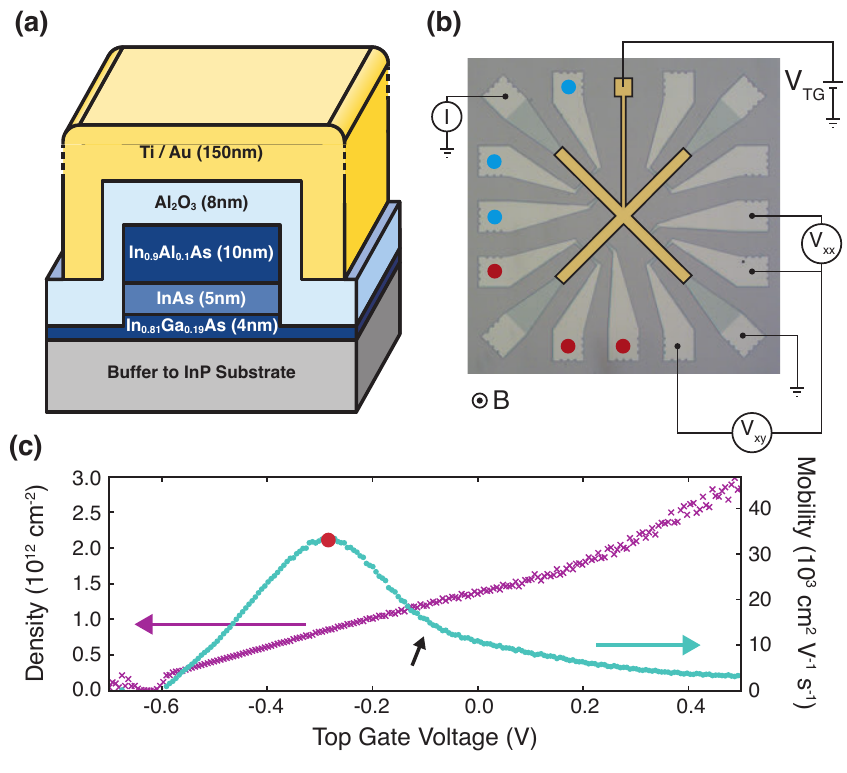}
\caption{\label{fig:fig1} (a) Cross section of a shallow InAlAs/InAs/InGaAs quantum well after removal of epitaxially grown Aluminum. A protective layer of \ce{Al2O3} is grown, along with a \SI{150}{\nano\meter} Ti/Au surface gate which is used to tune the electron density. (b) False color micrograph and schematic of the experimental setup showing one combination of current and voltage contacts. Red and blue dots indicate alternative measurement points. (c) The density (violet) and mobility (cyan) of sample B are extracted from magnetoconductance measurements as the top gate is swept. The red mark indicates the location of peak mobility. The black arrow indicates the location of the onset of second subband filling.}
\end{figure}

\section{\label{sec:exp}Experiment}

The devices are fabricated from a InAlAs/InAs/InGaAs quantum well grown \SI{10}{\nano\meter} below the surface on a 2'' (100) InP substrate \cite{manfra_hmob}. An \SI{8}{\nano\meter} Al layer is grown epitaxially on the surface of the heterostructure directly following the semiconductor growth. On each sample, a Hall bar geometry is defined using a dilute phosphoric acid etch, and Al is selectively removed over the Hall bar with an Aluminum wet etch (Transene type-D). Contact to the 2DEG is made using sections of un-etched Al, which forms an ohmic contact. The surface is then treated using either TMA as a reducing agent to remove the native oxide \cite{ingaas_redux,iiiv_cleanup} or with a ArH plasma to terminate charged impurity states \cite{BELL1998125}, and without breaking vacuum, a \SI{10}{\nano\meter} \ce{Al2O3} oxide is grown via ALD at \SI{200}{\celsius}, using a TMA precursor and either \ce{H2O} or \ce{O3} as an oxidizing agent. Finally, a \SI{150}{\nano\meter} Ti/Au gate is evaporated on the surface of the Hall bar to allow the electron density of the samples to be varied. Further details of the fabrication are contained in the Supplementary Information. A cross-sectional schematic of the Hall bar is given in Fig.~\ref{fig:fig1} (a). For each treatment/oxidizer pair, we fabricate two samples, one taken from within 1'' of the center of the wafer (denoted `near'), the other taken from the outer 1'' ring (denoted `far'), in order to account for variation in mobility as a function of distance from the center of the wafer \cite{watson_thesis}. A full list of tested sample parameters is given in Table~\ref{tab:sampparam}. We note that despite the higher quality of ALD oxides grown at higher temperatures, at temperatures higher than \SI{250}{\celsius}, diffusion of In and As occurs, and above \SI{300}{\celsius} \ce{In} begins to precipitate out of the substrate due to the desorption of \ce{As} \cite{PhysRevB.48.2807}.

\begin{table}
\caption{\label{tab:sampparam}%
A full listing of sample growth parameters that were tested. Two surface treatments (TMA Reduction and \ce{H2} Passivation) and two oxidizers (\ce{H2O} and \ce{O3}) are tested to find their effect on sample mobility. Each treatment and oxidizer pair are measured on two chips, one taken from the center of the growth wafer, the other from near the edge, in order to account for the effect of distance from the center of the wafer on mobility.}
\begin{ruledtabular}
\begin{tabular}{lll}
\textrm{Sample}&
\textrm{Treatment}&
\textrm{Precursor/Oxidizer}\\
\colrule
A & No Treatment & TMA/\ce{H2O} \\
B & TMA Reduction & TMA/\ce{H2O} \\
C & TMA Reduction & TMA/\ce{O3} \\
D & \ce{H2} Passivation & TMA/\ce{H2O} \\
E & \ce{H2} Passivation & TMA/\ce{O3} 
\end{tabular}
\end{ruledtabular}
\end{table}

Measurements were carried out in a dilution refrigerator with a base temperature of \SI{7}{\milli\kelvin}. A Cryo-CMOS based multiplexer is used to allow simultaneous measurement of up to 10 Hall bars in a single cool-down \cite{2019arXiv190807685P}. Magnetotransport measurements were performed to extract electron densities and mobilities using conventional AC lock-in techniques, with a \SI{10}{\nano\ampere} constant current. A representative Hall bar is shown in Fig.~\ref{fig:fig1} (b), with longitudinal ($R_{xx}$) and transverse ($R_{xy}$) resistance measured simultaneously. Three measurement points are defined around the edge of the Hall bar, indicated with blue, red and black dots, allowing multiple independent measurements of mobility and density to be made on each sample, from which statistics on each treatment are gathered. Hall bars were oriented at $\ang{45}$ to the $(011)$ and $(01\overline{1})$ plane to remove effects of any anisotropy along different crystallographic axes \cite{PhysRevB.67.045309,PhysRevB.77.235307}. For each sample, density and mobility are extracted as the top gate is swept. A representative measurement is shown for sample B, taken from near the center of the growth wafer, in Fig.~\ref{fig:fig1} (c). Mobility is extracted at \SI{0.05}{\tesla}, which is a sufficiently large offset to ensure we are no longer on the weak-antilocalization peak. For this sample, a peak value for mobility is extracted of \mob{33400} at a density of \den{8.73e11}, corresponding to a gate voltage of $V_{TG} = \SI{-0.27}{\volt}$, indicated by the red point in Fig.~\ref{fig:fig1} (c).

We can attribute different dominant scattering mechanisms to various ranges of the density \cite{Matsumoto_1974,PhysRevB.32.8126,PhysRevB.16.4446}. As the density increases from zero, scattering is predominantly caused by scattering off background impurities distributed through the heterostructure \cite{scattering}. Increased screening of impurities as the density is increased leads to an increase in mobility. As the gate voltage and density is further increased, the mobility is seen to peak, before reducing with increasing density. Increasing top gate voltage causes the the distribution of electrons in the quantum well to shift towards the surface \cite{doi:10.1063/1.119829,PhysRevB.93.235312}, and surface scattering becomes dominant over the increased impurity screening with higher density, leading to the decrease in mobility observed at higher densities.

\begin{figure}
    \includegraphics{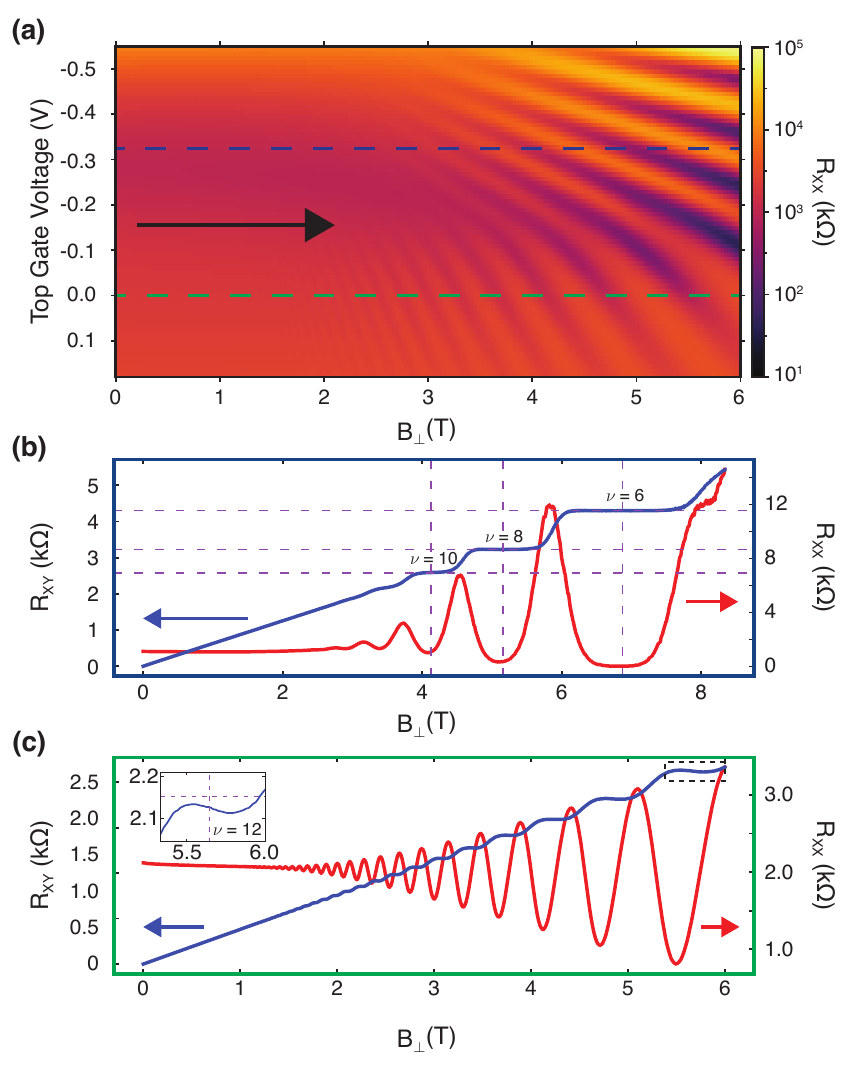}
    \caption{\label{fig:fig2}(a) The Landau fan for sample B. The black arrow marks the onset of second subband population, indicated by a change in slope of the Landau levels as a function of magnetic field and top gate voltage. The location of peak mobility is indicated by the blue dashed line. (b) Magnetoresistance taken at the point of highest mobility on sample B. Well resolved hall plateaus are observed, starting from $\nu = 10$ (see main text for details). (c) Magnetoresistance measurements of sample B to high field, taken at $V_{TG} = \SI{0}{\volt}$. Hall plateaus show an oscillation characteristic of parallel conduction paths. }
\end{figure}

The subband occupation can be extracted from magnetotransport measurements at high magnetic fields. Fig.~\ref{fig:fig2} (a), shows a Landau fan for sample B measured in a second cooldown, plotting $R_{xx}$ as $V_{TG}$ and $B$ are swept.  The onset of second subband population is marked by the arrow and occurs at $V_{TG} = \SI{-0.16}{\volt}$, denoted by the change in the slope of the location of Landau levels as a function of gate voltage and magnetic field \cite{PhysRevB.74.195313,STORMER1982707}. The onset of second subband population is also visible as a kink in the mobility as a function of $V_{TG}$, indicated by the black arrow in Fig.~\ref{fig:fig1} (c), caused by a reduction in the rate of filling of the first subband $N_{S1}$ relative to the total density $N_T = N_{S1} + N_{S2}$. 
 
The sample shows significantly different magnetotransport behavior when the first and second subbands are occupied. When the top gate is tuned to the value that maximizes mobility at low magnetic fields and only a single subband is occupied, well resolved Hall plateaus are observed from $\nu = 10$ onwards, with the Hall resistivity quantized to within $0.11\%$ of the theoretical value at $\nu = 6$, and a vanishing longitudinal resistance of $R_{xx} = \SI[per-mode=symbol]{2.4}{\ohm\per\sq}$. No Shubnikov-de Haas oscillations are observed. In contrast, when the second subband is occupied at $V_{TG} = \SI{0}{\volt}$ (Fig.~\ref{fig:fig2} (c)), clear Shubnikov-de Haas oscillations are visible from \SI{1.5}{\tesla} to \SI{6}{\tesla}, despite the much lower mobility of the sample at this point. This is caused by the increased screening of the impurity potential by the electrons of the second subband, and has been observed in measurements of low-mobility GaAs 2DEGs \cite{PhysRevB.38.7866}. Further evidence of second subband population is seen in the Hall plateaus which are not well quantized and exhibit an oscillatory behaviour (inset Fig.~\ref{fig:fig2} (c)), an effect attributed to parallel transport in the second subband.

\section{\label{sec:treat}Surface Treatments and Oxide Growth}

\begin{figure}
    \includegraphics[width=\linewidth]{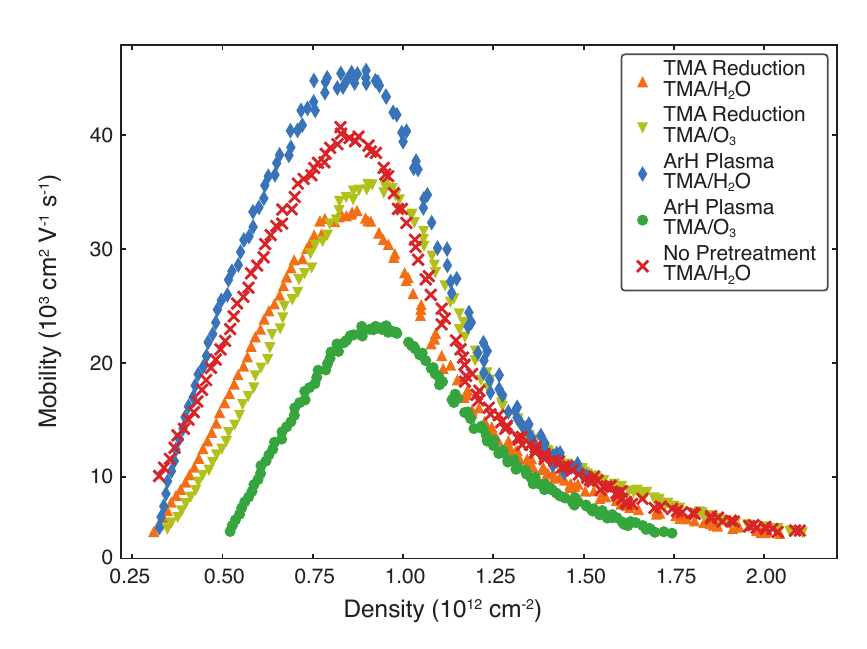}
    \caption{\label{fig:fig3} Representative mobility vs. density traces for each treatment, taken from samples near the center of the wafer. Samples oxidized with \ce{O3} show a shifted peak mobility relative to those oxidized with \ce{H2O}. A peak mobility of \mob{45300} is extracted for a sample treated with a Hydrogen plasma.}
\end{figure}

The native oxide layer in both GaAs and InAs is known to contain a large number of charged defects \cite{doi:10.1063/1.5054292,PhysRevB.49.11159}, caused by unpaired As atoms within the oxide formed by an excess of As during the oxidation of In and Ga \cite{doi:10.1063/1.3369540,Affentauschegg_2001}. These defects act as scattering sites at the surface of the wafer, and limit the mobility of samples above a certain density. Reducing the concentration of surface scattering sites through chemical treatment prior to dielectric deposition provides a clear pathway to increasing the sample mobility. 

The first approach that we examine is the removal of the native oxide through reduction by TMA\cite{doi:10.1063/1.3148723,Tallarida_2012,CLEVELAND2013167}. 
TMA is known to remove the surface oxides of InAs via the following reaction \cite{iiiv_cleanup}:
\begin{align}
    \ce{2Al(CH3)3 + In2O3 &-> 2In(CH3)3 + Al2O3} \\
    \ce{2Al(CH3)3 + As2O3 &-> 2As(CH3)3 + Al2O3}
\end{align}

For TMA treated samples, a \SI{1}{\second} pulse of TMA is applied to the surface in the ALD chamber, followed by a \SI{30}{\second} purge with \ce{N2} gas, at a \SI{200}{\celsius} process temperature. This pulse cycle is repeated 18 times to maximize the reaction time, prior to the growth of the dielectric.

The second approach that we examine is the removal of surface oxides and the passivation of charged impurities at the surface via the application of an ArH plasma to the chip \cite{CLEVELAND2013167,BELL1998125,doi:10.1116/1.586538}. For this process, a remotely generated ArH plasma is applied to the surface of the samples for a total of \SI{120}{\second} before the growth of the dielectric layer. Atomic hydrogen is known to bond to As atoms and saturate the dangling bonds, passivating the surface \cite{BELL1998125}. A hydrogen plasma is also known to selectively remove the surface oxide via dry etching, again leading to an abrupt semiconductor-dielectric interface \cite{doi:10.1063/1.92194,doi:10.1063/1.100961}.

\begin{figure}
\includegraphics[width=\linewidth]{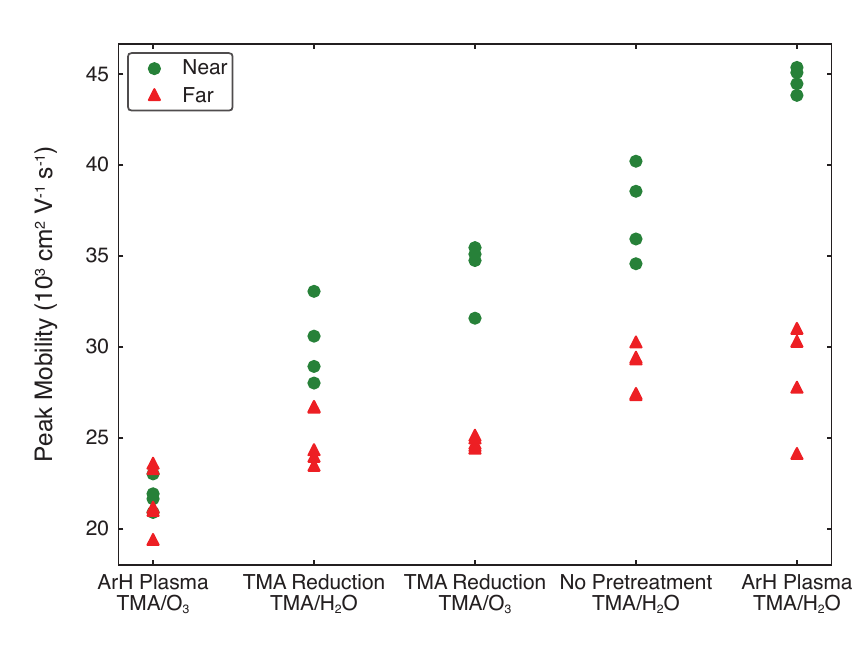}
\caption{\label{fig:fig4} Peak mobility achieved for different treatment and oxidizers. Measurements are made across different two samples, taken from near the center of the growth wafer (green) and from near the edge of the growth wafer (red), and at multiple locations on each Hall bar. Each point represents the peak mobility extracted from a sweep of gate voltage, as shown in red in Fig.~\ref{fig:fig1} (c), which were collected over multiple cooldowns, and at multiple measurement points.}
\end{figure}

In Fig.~\ref{fig:fig3} we plot mobility as a function of density for each treatment, with samples taken from near the center of the wafer. The use of ArH plasma in combination with oxide growth using TMA and \ce{H2O} as an oxidizer was found to increase the measured mobility relative to an untreated sample, showing the highest peak mobility for both near and far samples. In contrast, oversaturation with TMA causes a decrease in mobility compared to a single TMA exposure before \ce{Al2O3} growth. Finally, we note that the use of \ce{O3} as a precursor does not seem to be effective for the creation of a clean dielectric interface -- both ozone samples show a decreased quality relative to no treatment, however the peak mobility shifts to a higher density. Measurements are taken on both samples near the center of the growth wafer (near) and samples taken far from the center of the growth wafer (far), across multiple measurement points and multiple cooldowns, as shown in Fig.~\ref{fig:fig4}. While there exists a significant difference in the mobility of samples taken from different parts of the growth wafer, there remains a definite trend amongst similarly treated samples. The use of ArH plasma in combination with oxide growth using TMA and \ce{H2O} as an oxidizer results in a significant reduction in variance in mobilities for the highest quality samples.

\section{\label{sec:scat}Scattering Mechanisms}

\begin{figure}
    \includegraphics{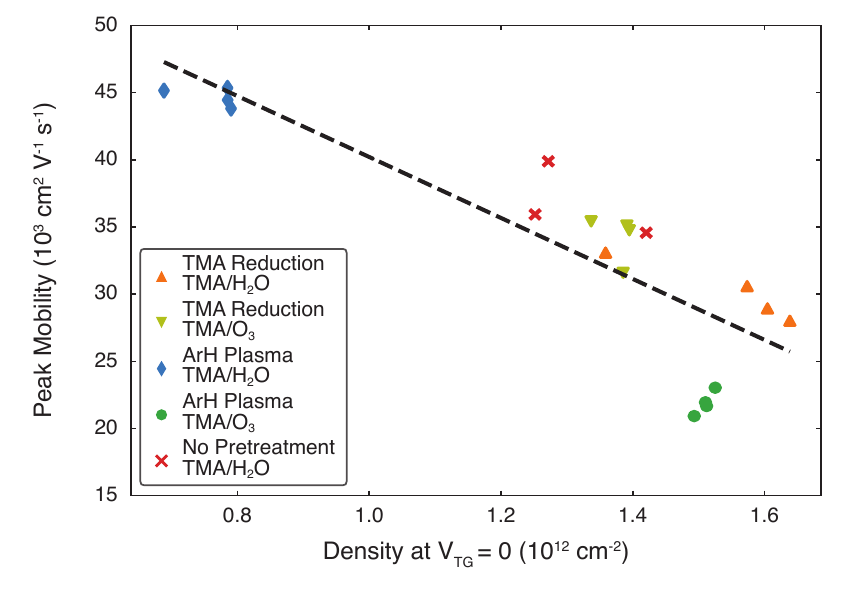}
    \caption{\label{fig:fig5}Scatter plot of density at zero gate voltage against the peak mobility, for samples taken near the center of the growth wafer. Samples with the lowest density at zero gate voltage have the highest measured peak mobility. The dashed black line is a linear fit, and is a guide to the eye.}
\end{figure}

Finally, we turn to a detailed examination of scattering mechanisms across different density ranges and surface treatments. Unlike semiconductors such as GaAs where the Fermi level is pinned in the band gap, the location of the Fermi level in InAs has been shown to depend sensitively on surface states. Even in nominally undoped heterostructures, an electron density in the quantum well at zero gate voltage is induced by charged impurities at the surface \cite{PhysRevLett.66.2243,Affentauschegg_2001}. As the concentration of charged impurities is decreased, the electron density in the quantum well at zero gate voltage is decreased towards zero. Fig.~\ref{fig:fig5} shows the density at zero gate voltage against the peak mobility. We observe an inverse relationship, with the samples that have the highest mobility also having the lowest intrinsic electron density.  Atomic hydrogen plasma is thus an effective method for terminating these charged impurities prior to dielectric growth. 

In contrast, samples treated with TMA see either no significant change in the density of charged surface states or see an increase relative to no pre-treatment. Although TMA treatment has been demonstrated to be effective in removing the surface oxide, such studies largely investigate the optical properties of the cleaned surface, using either X-ray photo-emission spectroscopy \cite{doi:10.1063/1.3148723,Tallarida_2012,CLEVELAND2013167} or infra-red spectroscopy \cite{doi:10.1021/jp412455y}, rather than the electrical properties. We suggest that the inconsistency between previous studies and our result can be explained by the growth of an \ce{Al2O3} layer. This acts as a diffusion barrier at \SI{200}{\celsius} and terminates the native oxide removal process before completion \cite{HENEGAR2016870}, which in this case limits the effectiveness of the treatment.




To understand the reduced mobility observed when using ozone as the oxidizer in the ALD process, we examine the relationship between density and mobility when peak mobility is achieved (See supplementary for additional data). The peak mobility in \ce{O3} samples is shifted towards higher densities compared to those samples that use \ce{H2O}. Previous studies have found that the \ce{AlO_x} grown by ozone is oxygen-rich relative to the optimal stoichiometry for aluminum oxide \cite{ingaas_redux,10.1021/cm0608903}. The increased incorporation of oxygen in the oxide will appear as remote charged impurity scatterers distributed throughout the dielectric \cite{scattering}, and therefore a higher electron density has to be reached before these are fully screened.

\section{Conclusion}
In summary, we find scattering off charged surface impurities at the \ce{Al2O3} interface is a limiting factor in mobility in the current generation of shallow InAs quantum wells. For a quantum well \SI{10}{\nano\meter} from the surface, the application of an ArH plasma prior to dielectric growth is effective in increasing the peak mobility to $\sim \mob{45300}$, and reduces the variance in mobility compared to untreated samples or samples exposed to TMA saturation. For all samples, we find that the second subband is occupied at $V_{TG} = \SI{0}{\volt}$. This is a complicating factor in the search for MZM in InAs 2DEGs, as an odd number of filled subbands are then required for their formation  \cite{s41578-018-0003-1}. For the current generation of samples, this condition can only be met with a single populated subband whilst achieving sufficiently high mobility. Due to pinning of the Fermi level in InAs, we find that a significant negative gate voltage will be necessary to tune into this regime.

\begin{acknowledgments}
This research was supported by the Microsoft Corporation and the Australian Research Council Centre of Excellence for Engineered Quantum Systems (EQUS, CE170100009). The authors acknowledge the facilities as well as the scientific and technical assistance of the Research \& Prototype Foundry Core Research Facility at the University of Sydney, part of the Australian National Fabrication Facility.
\end{acknowledgments}


\bibliography{surface}

\end{document}